\documentclass[sigconf,natbib=true,nonacm]{acmart}
\usepackage{color}
\usepackage{graphicx}
\usepackage{xspace}
\usepackage{epsfig}
\usepackage{url}
\usepackage{booktabs}
\usepackage{amsmath}
\usepackage{balance}

\author{Haya Nachimovsky}
\email{haya.nac@campus.technion.ac.il}
\affiliation{%
  \institution{Technion}
  \country{Israel}
  }

\author{Moshe Tennenholtz}
\email{moshet@technion.ac.il}
\affiliation{%
  \institution{Technion}
  \country{Israel}}

\author{Fiana Raiber}
\email{fiana@yahooinc.com}
\affiliation{%
  \institution{Yahoo Research}
  \country{Israel}}

\author{Oren Kurland}
\email{kurland@technion.ac.il}
\affiliation{%
  \institution{Technion}
  \country{Israel}}

\usepackage{subcaption}
\usepackage{bbm}
\usepackage{algorithm}
\usepackage{algorithmic}
\usepackage{amsmath}
\usepackage{pifont}

\newcommand{\haya}[1]{\textcolor{magenta}{Comment: #1}}

\newcommand{\cmark}{\ding{51}}
\newcommand{\xmark}{\ding{55}}

\newcommand{\floor}[1]{\left\lfloor #1 \right\rfloor}
\newcommand{\ceil}[1]{\left\lceil #1 \right\rceil}
\newcommand{\median}{\operatorname{median}}

\newcommand{\sAlg}{\(s_{\mathrm{alg}}\) }

\newcommand{\LTRNoX}{LTR}
\newcommand{\semNoX}{NEU}
\newcommand{\llmNoX}{AI}
\newcommand{\nollmNoX}{$\urcorner$\llmNoX}
\newcommand{\LTR}{\LTRNoX\xspace}
\newcommand{\sem}{\semNoX\xspace}
\newcommand{\llm}{\llmNoX\xspace}
\newcommand{\nollm}{\nollmNoX\xspace}

\newcommand{\A}{\LTRNoX$\wedge$\llmNoX\xspace}
\newcommand{\B}{\LTRNoX$\wedge\urcorner$\llmNoX\xspace}
\newcommand{\C}{\semNoX$\wedge$\llmNoX\xspace}
\newcommand{\D}{\semNoX$\wedge\urcorner$\llmNoX\xspace}

\newcommand{\nimrodBestShort}{Raifer}
\newcommand{\allPredFeatures}{All}
\newcommand{\docsPercentage}{Docs}
\newcommand{\avgRank}{Rank}
\newcommand{\OneWin}{$x=1$}
\newcommand{\TwoWins}{$x=2$}
\newcommand{\ThreeWins}{$x=3$}

\newcommand{\lambdamart}{LambdaMART\xspace}
\newcommand{\bert}{BERT\xspace}

\newcommand{\fracStop}{StopwordsRatio\xspace}
\newcommand{\stopCover}{StopwordsCover\xspace}

\newcommand{\llewq}{$L \le W_q$\xspace}
\newcommand{\lgwq}{$L > W_q$\xspace}
\newcommand{\wt}{$W_t$\xspace}
\newcommand{\wq}{$W_q$\xspace}

\newcommand{\llewqf}{$\mathbf{L \le W_q}$\xspace}
\newcommand{\lgwqf}{$\mathbf{L > W_q}$\xspace}
\newcommand{\wtf}{$\mathbf{W_t}$\xspace}
\newcommand{\wqf}{$\mathbf{W_q}$\xspace}

\newcommand{\winnerPrevCentroid}{PWC\xspace}

\newcommand{\docr}[1]{D_{#1}}
\newcommand{\winr}[1]{W^{\query}_{#1}}
\newcommand{\allwinr}[1]{W^{Q}_{#1}}

\newcommand{\centroidMacro}{TopicMacro\xspace}
\newcommand{\cmOne}{SIM(\docNow,\winnerPrevCentroid)\xspace}
\newcommand{\cmTwo}{SIM(\docPrev,\winnerPrevCentroid)\xspace}

\newcommand{\QueryPrev}{QueryRank\xspace}
\newcommand{\qpOne}{BEST\xspace}
\newcommand{\qpTwo}{MEDIAN\xspace}
\newcommand{\qpThree}{WORST\xspace}

\newcommand{\PrevImprove}{PrevChange\xspace}

\newcommand{\PrevPosition}{TopicRank\xspace}
\newcommand{\ppOne}{BestPrevRank\xspace}
\newcommand{\ppTwo}{MedianPrevRank\xspace}
\newcommand{\ppThree}{WorstPrevRank\xspace}

\newcommand{\docLen}{Len\xspace}

\newcommand{\qmajority}{QueryMajority\xspace}
\newcommand{\qmajorityShort}{QMaj\xspace}
\newcommand{\tmajority}{TopicMajority\xspace}
\newcommand{\tmajorityShort}{TMaj\xspace}

\newcommand{\nimrodComp}{Raifer et al. \cite{Raifer+al:17a} \xspace}

\newcommand{\bm}{BM25\xspace}
\newcommand{\lmir}{LMIR\xspace}
\newcommand{\tf}{TF\xspace}
\newcommand{\normtf}{NormTF\xspace}
\newcommand{\len}{LEN\xspace}
\newcommand{\fracstop}{FracStop\xspace}
\newcommand{\stopcover}{StopCover\xspace}
\newcommand{\ENT}{ENT\xspace}

\newcommand{\querycover}{QueryTermsCover\xspace}
\newcommand{\lmirjm}{LMIR.JM\xspace}
\newcommand{\sumtf}{SumTF\xspace}

\newcommand{\sumtfidf}{SumTF-IDF\xspace}

\newcommand{\MABOT}{MABOT\xspace}
\newcommand{\MTBOT}{MTBOT\xspace}
\newcommand{\MSBOT}{MSBOT\xspace}
\newcommand{\NMABOT}{NMABOT\xspace}
\newcommand{\NMTBOT}{NMTBOT\xspace}
\newcommand{\NMSBOT}{NMSBOT\xspace}

\newcommand{\obsA}{\emph{full}\xspace}
\newcommand{\obsT}{\emph{partial}\xspace}
\newcommand{\obsS}{\emph{self}\xspace}

\newcommand{\markov}{\emph{Markov}\xspace}
\newcommand{\nonmarkov}{\emph{full}\xspace}

\newcommand{\myparagraph}[1]{\vspace{0.6\baselineskip}\noindent{\textbf{#1}}.~}

\newcommand{\omt}[1]{}
\newcommand{\firstmention}[1]{{\bf #1}}

\newcommand{\round}{l}

\newcommand{\query}{q}

\newcommand{\arbTerm}{t}
\newcommand{\termsSet}{\mathcal{S}}

\newcommand{\set}[1]{\{#1\}}
\newcommand{\definedas}{\stackrel{def}{=}}

\newcommand{\ltr}{LTR\xspace}
\newcommand{\ourAlg}{\rforest}
\newcommand{\psvm}{PSVM\xspace}
\newcommand{\lsvm}{LSVM\xspace}
\newcommand{\lreg}{LReg\xspace}
\newcommand{\rforest}{RForest\xspace}

\newcommand{\FOne}{F1\xspace}
\newcommand{\acc}{Acc\xspace}

\newcommand{\ent}{Entropy\xspace}

\newcommand{\simToInit}{SimInit\xspace}

\newcommand{\metaFt}{Macro\xspace}

\newcommand{\atomicFt}{Micro\xspace}

\newcommand{\queryFt}{Query\xspace}

\newcommand{\stopwordsFt}{Stopwords\xspace}

\newcommand{\noQueryStopsFt}{$\urcorner$Query$\urcorner$Stopwords\xspace}
\newcommand{\noQueryStopsFtNoX}{$\urcorner$Query$\urcorner$Stopwords}
\newcommand{\docNow}{D\xspace}
\newcommand{\docPrev}{PD\xspace}
\newcommand{\winnerPrev}{PW\xspace}
\newcommand{\fOne}{SIM(\docNow,\docPrev)\xspace}
\newcommand{\fTwo}{SIM(\docNow,\winnerPrev)\xspace}
\newcommand{\fThree}{SIM(\docPrev,\winnerPrev)\xspace}
\newcommand{\addOne}{ADD(\winnerPrev)\xspace}
\newcommand{\addTwo}{ADD($\urcorner$\winnerPrev)\xspace}
\newcommand{\rmOne}{RMV(PW)\xspace}
\newcommand{\rmTwo}{RMV($\urcorner$PW)\xspace}

\newcommand{\llew}{L$\le$W\xspace}
\newcommand{\lgw}{L$>$W\xspace}

\newcommand{\awinner}{AllWinners\xspace}
\newcommand{\awinnerShort}{AllW\xspace}
\newcommand{\aloser}{AllLosers\xspace}
\newcommand{\aloserShort}{AllL\xspace}
\newcommand{\random}{Random\xspace}
\newcommand{\randomShort}{Rand\xspace}

\title{Competitive Retrieval: Going Beyond the Single Query}

\begin{abstract}
  Previous work on the competitive retrieval setting focused on a
  single-query setting: document authors manipulate their documents so
  as to improve their future ranking for a given query. We study a
  competitive setting where authors opt to improve their document's ranking for
  multiple queries. We use game theoretic analysis to
  prove that equilibrium does not necessarily exist. We then empirically show that it is more
  difficult for authors to improve their documents' rankings for
  multiple queries with a neural ranker than with a state-of-the-art
  feature-based ranker. We also present an effective approach for predicting the document most highly ranked in the next induced ranking.
\end{abstract}

\begin{document}
\maketitle

\section{Introduction}
\label{sec:intro}
In the Web search setting, publishers (authors) of documents are
sometimes incentivized to have their documents highly ranked by search
engines for some queries. The basic motivation is the fact that
documents at top ranks often attract most user engagement
\cite{Joachims+al:05a}. As a result of their ranking incentives, publishers might
modify their documents in response to rankings induced by a search
engine so as to improve their future rankings. The search setting then
becomes competitive \cite{Kurland+Tennenholtz:22a} with corpus
dynamics driven in part by ranking-incentivized document
modifications. These modifications are often referred to as search
engine optimization (SEO) \cite{Gyongyi+Molina:05a}.


Lately, there has been a growing body of work on studying
ranking-incentivized document modifications, and more generally, on
competitive retrieval
\cite{Basat+al:17a,Raifer+al:17a,Goren+al:18a,Goren+al:20,Goren+al:21a,Liu+al:22a,Song+al:22a,Wang+al:22a,Wu+al:22a,Chen+al:23a,Chen+al:23b,Liu+al:23a,Vasilisky+al:23,Wu+al:23a}. This
renewed interest in the often dubbed {\em adversarial retrieval}
setting \cite{Castillo+Davison:10a} can be attributed in part to a
large volume of work on adversarial machine learning in the era of 
neural network models
\cite{Szegedy+al:14a,Huang+al:17a,Jia+Liang:17a,Papernot+al:17a,Xie+al:17a,Kreuk+al:18a,Jia+al:19a,Zhang+al:19a}.

The focus of some recent work on competitive retrieval is on
algorithmic attacks on ranking functions; e.g., substituting terms
with their synonyms
\cite{Goren+al:21a,Liu+al:22a,Song+al:22a,Wang+al:22a,Wu+al:22a,Chen+al:23a,Chen+al:23b,Liu+al:23a}. There
is also an emerging thread of work on analyzing, and improving, the
robustness of ranking functions to ranking-incentivized document
modifications
\cite{Basat+al:17a,Goren+al:18a,Chen+al:23a,Vasilisky+al:23,Wu+al:23a}. Another
type of work, which is our focus in this paper, is analyzing human
strategies of ranking-incentivized document manipulations
\cite{Raifer+al:17a} and their resultant effect on the corpus
\cite{Basat+al:17a,Raifer+al:17a,Goren+al:21a}. For example, it was
shown theoretically and empirically that a prominent (worthwhile)
strategy of publishers is to mimic content in documents that were
highly ranked in the past for queries of interest
\cite{Raifer+al:17a}, which results in an {\em herding} effect \cite{Goren+al:21a}.

Almost all recent work on competitive retrieval has focused on a
single-query setting; i.e., assuming a publisher
modifies her document to improve its ranking for a single
query\footnote{We discuss in Section \ref{sec:rel} the two exceptions:
  algorithmic attacks for topically-related queries \cite{Liu+al:23a}
  and the corpus effect of modifying documents for two
  queries (topics) \cite{Basat+al:17a}. Algorithmic attacks are outside the
  scope of this paper.}. However, publishers often opt to have
their document highly ranked for multiple queries; e.g., those representing a topic in the document. Accordingly, in this paper we present the first --- to the best of our
knowledge --- theoretical and empirical study of ranking-incentivized
document-modification strategies of (human) publishers opting to
promote their documents for multiple queries.

Using game theoretic analysis we show that when publishers modify
their documents  to improve their rankings for {\em multiple} queries, there
is not necessarily an equilibrium. This result implies instability: 
a potentially endless document modification race which can have negative effects on users; e.g., documents at top ranks consistently change not due to pure editorial considerations but rather due to ranking incentives. This theoretical result is in contrast
to recent findings about the existence of an equilibrium when
documents are modified for a {\em single} query
\cite{Raifer+al:17a}.

We then fully characterize
the conditions for an equilibrium in the multiple-queries setting for a basic family of ranking
functions. Furthermore, we show that {\em best response dynamics}
\cite{Nisan+al:07a} does not necessarily converge to an equilibrium in
case it exists. Best response dynamics is a process where each
publisher modifies her document to attain the best possible ranking 
given the documents written by all other
publishers. The implication is that reaching a steady state in case it is achievable might call
for an external intervention; e.g., of the search engine.

We next present an empirical study of document manipulation
strategies applied by publishers that opt for improved ranking for
multiple queries representing the same information need (topic). The
study is based on ranking competitions we organized between
students. Our competitions were inspired by those
organized for the single-query setting
\cite{Raifer+al:17a,Goren+al:18a,Goren+al:21a}. The competitions we
report differ from previous competitions
\cite{Raifer+al:17a,Goren+al:18a,Goren+al:21a} not only by the virtue
of having publishers modify their documents for multiple queries
rather than a single one, but also by two additional aspects. First,
our competitions are the first to employ not only a feature-based
learning-to-rank (LTR) method \cite{Liu:11a} as the ranking function, but also a
neural ranker \cite{Lin+al:20a}. Second, in some of our competitions, we encouraged the
participants to use generative AI tools (e.g., GPT \cite{gpt23}) to help modify
their documents. The competitions were
approved by ethics committees. The data of the competitions and our code will be made public upon publication of this paper.

Analysis of the ranking competitions revealed that, as is the case for the single-query setting
\cite{Raifer+al:17a}, publishers tend to mimic content from documents
that were previously highly ranked. This trend was more prominent when publishers were not allowed to use generative AI tools. We also found that the neural ranker's rankings
for different queries representing the same information need were more diverse than those of the LTR method. Accordingly, it was much harder for publishers to have their documents highly ranked for multiple
queries with the neural ranker. 

Finally, as in work on the single-query setting \cite{Raifer+al:17a},
we pursue the task of predicting which document among those not ranked
first in the last ranking will become the most highly ranked assuming that the top
ranked document is going to change. As it turns out, utilizing
information induced from rankings for other queries representing the
same information need can help improve prediction effectiveness for the query at hand. This is yet another difference between the multiple-queries setting and the single-query setting.

Our contributions can be summarized as follows:
\begin{itemize}
\item A theoretical and empirical study of document manipulations intended to improve ranking for {\em multiple} queries.
  \item A full game theoretic characterization of when
    equilibrium of document manipulation strategies exists.
    \item Showing that best response dynamics might not converge even when an equilibrium exists.
    \item Organizing novel ranking competitions where publishers
      compete for multiple queries. Additional novel aspects are using a neural ranker and allowing publishers to use generative AI tools for document modification.
      \item A method for predicting which documents among those not ranked first might be ranked first in the next ranking; and, demonstrating the merit of using information induced from other queries to this end.
  \end{itemize}

\section{Related Work}
\label{sec:rel}

The work most related to ours is based on a game theoretic and empirical
analysis of ranking incentivized document manipulations applied for a {\em single} query
\cite{Raifer+al:17a}. In contrast to our {\em multiple} queries settings, an
equilibrium in the single-query setting always exists. Raifer et
al. \cite{Raifer+al:17a} and Goren et al. \cite{Goren+al:21a} found using ranking competitions that
publishers tend to mimic content from documents highly ranked in the
past for a query: a strategy justified by Raifer et al. \cite{Raifer+al:17a} using a game
theoretic analysis. We found a similar pattern in ranking
competitions organized for the multiple-queries setting. We use both a neural
ranker and a feature-based ranker while Raifer et
al. \cite{Raifer+al:17a} used only the latter. In contrast to our
work, Raifer et al. \cite{Raifer+al:17a} did not allow the use of
generative AI tools for document manipulation in their ranking
competitions.

Several algorithmic attacks on neural retrieval methods have been
recently reported
\cite{Goren+al:21a,Liu+al:22a,Song+al:22a,Wang+al:22a,Wu+al:22a,Chen+al:23a,Chen+al:23b,Liu+al:23a}. The
attacks are almost always for a single query, except for
an attack for topically related queries
\cite{Liu+al:23a}. In contrast, in our game theoretic analysis, the
document manipulations are changes of the relative emphasis of a
document on different queries. This type of strategic manipulation has not been studied in previous work to the best of our knowledge. Furthermore, we are not aware of any other studies, as ours, of human document-modification strategies for multiple queries with or without generative AI tools.


The suboptimality of the probability ranking principle
\cite{Robertson:77a} in competitive retrieval settings was
demonstrated using a game theoretic approach \cite{Basat+al:17a}. Publishers can write either a single topic document or a two-topics
document with equal emphasis for the topics. Since the resulting
ranking game has a finite number of players (publishers) and a finite
number of actions a player can play (i.e., documents they can write),
by Nash's theorem \cite{Nash1950}, a mixed-strategies
equilibrium exists. That is, documents are stochastically created by
applying a distribution over a document set. The setting is less
realistic than the one we address here where publishers can write a
document with different emphasis (represented as a real number) on
different queries. Our resulting games have an infinite number of
actions a publisher can take, and we show that there is not
necessarily an equilibrium.



\section{Game Theoretic Analysis}
\label{sec:model}

To analyze the search setting when
publishers modify their documents to have them highly ranked for
multiple queries, we use game theory. We consider
publishers as players in a ranking game: the documents they publish
are their strategies and the search engine's ranking function is the
mediator \cite{Kurland+Tennenholtz:22a}.

As noted above, some previous work used game theory to analyze the single-query search setting where
publishers opt to promote their documents for a single
query \cite{Raifer+al:17a}; single-peaked ranking
functions were used \cite{Raifer+al:17a}. Several sparse retrieval methods
are single peaked; e.g., cosine between tf.idf vectors and negative KL
divergence over multinomial
distributions \cite{Zhai+Lafferty:05a}. Furthermore, considering the common technique of
stuffing query terms in the document to improve its ranking \cite{Gyongyi+Molina:05a}, the
retrieval function could also be viewed as single peaked with respect to query term occurrence. That is, moderately
increasing query term occurrence usually results in monotonic increase of retrieval
scores. However, as from a certain point, further increasing query term occurrence can result in increased penalty to retrieval scores when document quality measures are used \cite{Bendersky+al:11b}. There is also some empirical
evidence in prior work \cite{Raifer+al:17a} and ours (Section \ref{sec:eval}) that publishers might view the undisclosed ranking function
as single peaked. When considering incomplete information about the
ranker (unknown peak), the ranking game has a minimax regret equilibrium
where publishers mimic content of documents that were previously the
highest ranked \cite{Raifer+al:17a}.

Turning to our multiple-queries setting, we assume that a document has split emphasis on various queries; e.g., different passages match different queries to varying degrees. We assume as Raifer et al. \cite{Raifer+al:17a} a single peaked ranking function, and focus as Basat et al. \cite{Basat+al:15a} on
ranking games where the retrieval function is disclosed (i.e., its
peak is known).  We show that the
resulting situation in the multiple-queries setting with a disclosed ranking function is considerably more subtle than that
in the single query setting where it is trivial to show that an equilibrium exists. Specifically, we prove that there is not
necessarily an equilibrium in the multiple-query setting and fully characterize the cases when it exists.

\subsection{Game Definition}
\label{sec:gameDef}
Let $G =\langle n, m, p \rangle$ be a ranking game defined as follows:
$n$ publishers (players) write and modify documents in a corpus $D$; a query set $Q$
contains $m$ queries: $\{q_1,\ldots,q_m\}$. Each player writes a single document $d \in D
\definedas \{(d^1,\ldots,d^m) \in[0,1]^m : \sum_{j=1}^m d^j \le 1\}$, where $d^j$
($\in [0,1]$) is the relative emphasis of the document $d$ on information
relevant to query $q_j$.\footnote{In reference to work on focused
  retrieval \cite{Geva+al:11a}, $d^j$ can be defined as the portion of
  the text in $d$ that is marked as relevant to query $q_j$.}

Let $f: [0,1] \rightarrow [0,1]$ be a single peak function: there exists a unique $p \in [0,1]$ such that $f$ is
monotonically increasing in $[0,p]$ and monotonically decreasing in
$[p,1]$.  The retrieval function $r: D \times Q \rightarrow [0,1]$
assigns document $d~(\in D$) the retrieval score $r(d, q_j) \definedas f(d^j)$ with respect to query $q_j$ ($j \in \{1,\ldots,m\}$);
documents are ranked in descending order of scores
with ties broken arbitrarily.

We will refer to document $d_i$ written by player $i$ as the
(pure) {\bf strategy} of player $i$ in the game\footnote{There
  is also a notion of mixed strategies where players apply a
  distribution over pure strategies. Herein, we focus on pure
  strategies.}. That is, the strategy $d_i$ is defined by the choice of
relative emphasis on each query in the document $d_i$. A {\bf strategy profile}
$s \definedas (d_1,\ldots,d_n)$ is the tuple of pure strategies of the players in
the game; i.e., the set of documents in the corpus.

The {\bf utility} of a player $i$ who wrote document $d_i$, denoted
$U_i(s)$, is the number of queries for which $d_i$ is ranked
first; $d_i$'s rank for a query depends on the
strategies of all other players (i.e., their documents). To summarize, $G$ is a strategic game where $n$ players
(publishers) opt to maximize their documents' ranks for $m$  queries.

A strategy profile $s$ is a (pure) {\bf Nash equilibrium} if there is no incentive for any player $i$ to change her strategy given the strategies $\{d_j\}_{j \ne i}$ of all other players; i.e., a change will result in reduced utility.

\subsection{Game Analysis}
\label{sec:gameAnalysis}
Our goal is to find a full characterization of the
existence (or lack thereof) of pure Nash equilibria for $G$, given $n$, $m$ and $p$.

It is easy to see that for a small enough $p$, there exists a pure Nash equilibrium where the documents published by all players have the same emphasis on all queries:\footnote{Recall that the retrieval function has a single peak at $p$.}
\begin{lemma}
    \label{lem: small-p}
    If $p \le \frac{1}{m}$, then the profile where all players write $d = (p, \ldots, p)$ is a pure Nash equilibrium.
\end{lemma}

To extend our analysis for any $p$ ($\in [0,1]$), we start by
analyzing the case of two players ($n=2$).  This case has an interesting property, because
in any pure Nash equilibrium, the two players have the same utility.
They can attain this utility, for example, by using the same strategy (i.e., writing the same document):
\begin{lemma}
    \label{lem:copycat-strategy}
    Let $n=2$ and $m>n$.
    If $s=(d_1,d_2)$ is a pure Nash equilibrium, then $U_1(s)=U_2(s)=\frac{m}{2}$.
\end{lemma}

\begin{proof}
    Observe that for any profile $s=(d_1,d_2)$, $U_1(s)+U_2(s)=m$.
    Assume by contradiction that $U_1(s) \neq U_2(s)$, w.l.o.g.\ $U_1(s) < U_2(s)$; i.e., $U_1(s) < \frac{m}{2} < U_2(s)$.
    If player 1 adopts player 2's strategy, i.e., writes $d_1'=d_2$ then $U_1(d_1',d_2)=U_2(d_1',d_2) = \frac{m}{2} > U_1(d_1,d_2)$; hence, $s=(d_1,d_2)$ is not a pure Nash equilibrium --- a contradiction.

\end{proof}

We use Lemma \ref{lem:copycat-strategy} to fully characterize in Theorem \ref{thm:two-players} the
existence of pure Nash equilibrium for $G$ when $n=2$ (i.e., two players).

\begin{theorem}
\label{thm:two-players}
    If $n=2$ and $m>n$, then the game $G$ has a pure Nash equilibrium iff $p \le \frac{1}{m-1}$.
\end{theorem}
\begin{proof}
    We only provide a proof sketch due to space considerations and as it is technical. 
  Say a strategy profile $(d_1, d_2)$ is a pure Nash equilibrium; let $d_2 = (d_2^1,\ldots d_2^m)$, and assume w.l.o.g.\ that $d_2^1 \ge d_2^2 \ge \ldots \ge d_2^m$. We show that there cannot be two queries $q_i$ and $q_j$ such that $d_2^{i} < p$ and $d_2^{j} < p$. This allows to show that if there is an equilibrium, then $p \le \frac{1}{m-1}$. If $p \le \frac{1}{m}$, then by Lemma~\ref{lem: small-p} the profile where both players publish $d = (p, \ldots, p)$ is a pure Nash equilibrium. We show that for $\frac{1}{m} < p \le \frac{1}{m-1}$ the profile where $d_1 = d_2 = (p, \ldots, p, 1-p\cdot(m-1))$ is a pure Nash equilibrium.
\end{proof}

To extend the analysis to a game with more than two players, we start by considering the case of $m>n$:
\begin{theorem}
\label{thm: more-queries-than-players}
    The game $G=\langle n, m, p \rangle$ with $n<m$ has a pure Nash equilibrium iff $p \le \frac{1}{\ceil{\frac{2 \cdot m}{n} - 1}}$.
\end{theorem}
\begin{proof}
The central idea in the proof is to show that in any pure Nash equilibrium the following properties must hold:
(i) the highest ranked document $d$ for each query $q_j$ has $d^j \ge p$; and,
(ii) for each player there is at most one query for which her document is ranked first solely.
The proofs of these properties are mostly technical and are omitted due to space considerations.
For any $p > \frac{1}{\ceil{\frac{2 \cdot m}{n} - 1}}$, these properties cannot hold, and therefore there is no pure Nash equilibrium.
Conversely, we show that if $p \le \frac{1}{\ceil{\frac{2 \cdot m}{n} - 1}}$ then we can construct a pure Nash equilibrium as follows.
  
Given some strategy profile $s = (d_1, \ldots, d_n)$, let $h_j(s)$ denote the number of
documents assigned the same highest retrieval score for query $q_j$ when $s$
is played. The strategy profile \sAlg obtained by the
following algorithm is a pure Nash equilibrium (proof omitted due to space considerations):
\begin{algorithm}[H]
\caption{Equilibrium construction for $G=\langle n, m, p \rangle$ with $n<m$}
\begin{algorithmic}
  \footnotesize
    \STATE Initialize $d_i^j=0$ for all $i,j$.
    \STATE $k = \floor{\frac{1}{p}}$
    \FOR{$t = 1$ \TO $k$}
        \FOR{$i = 1$ \TO $n$}
            \STATE $j^{*} = \arg\min_{j \in \{1,\ldots, m\}} h_j(d_1, \ldots, d_n)$
            \STATE $d_i^{j^{*}} = p$
        \ENDFOR
    \ENDFOR
\end{algorithmic}
\label{alg:build-equilibrium}
\end{algorithm}


\end{proof}

In the case $n\ge m$, it is easy to see that a pure Nash equilibrium exists for every $p$.

Specifically, w.l.o.g. each (disjoint) group of $\floor{\frac{n}{m}}$ players is
assigned to an arbitrarily different query $q_j$ among $q_1,\ldots,q_m$; each player in this group
writes a document $d$ with $d^j=p$ and $d^k=0$ for $k \ne j$. Together with Theorem \ref{thm:
  more-queries-than-players}, we arrive to the final result:
\begin{corollary}
    G=$\langle n, m, p \rangle$ has a pure Nash equilibrium iff $p \le \frac{1}{\max{\{\ceil{\frac{2 \cdot m}{n} - 1}, 1\}}}$.
\end{corollary}

\subsection{Best response dynamics}
The {\bf best response} of a player to the strategies played by all other
players (i.e., their documents) is the strategy (document) that
maximizes her utility. Iteratively finding best responses until
convergence (if exists) --- a process known as {\bf best response dynamics} --- is a standard mechanism for reaching Nash
equilibria, specifically in {\em potential games} \cite{Nisan+al:07a}. In these games, the best response dynamics is guaranteed to converge to an equilibrium. We now show that in
our ranking games the best response dynamics does not
necessarily converge. Hence, these are not potential games. An
important implication is that in order to achieve stability
(specifically, equilibrium), the game might need external intervention.

Consider the game $G=\langle n, m, p \rangle$ for $n=2, m=3$ and
$p=\frac{1}{2}$. We show that the best response dynamics does not
converge in this game, although by Theorem~\ref{thm:two-players} an
equilibrium exists. Let $s_t$ denote the strategy profile played at round $t$ by two players. The following sequence of strategies is a best response dynamics given an initial strategy profile $s_0 = \left((0.3, 0.4, 0), (0.2, 0.3, 0.5) \right)$: $s_1 =
\left((0.3, 0.4, 0), (0.4, 0.5, 0.1) \right)$, $s_2 = \left((0.5, 0.3,
0.2), (0.4, 0.5, 0.1) \right)$, $s_3 = \left((0.5, 0.3, 0.2), (0, 0.4,
0.3) \right)$. For example, since player $1$'s document at round $0$ is
$(0.3,0.4,0)$ then player 2's document at round $1$, $(0.4,0.5,0.1)$,
attains the maximal possible utility (3) as it outranks player $1$'s
document for all three queries. Player $1$'s best possible utility given player $2$'s document from round 1, $(0.4,0.5,0.1)$, is 2, which is obtained by writing, for example, a document $(0.5,0.3,0.2)$ at round $2$, and so forth. Since $s_0$ and $s_3$ are symmetric, the dynamics does not converge.


\newcommand{\figWidth}{1.7in}
\newcommand{\figWidthSecond}{1.9in}
\newcommand{\figHeight}{1.2in}
\newcommand{\figHeightSecond}{1.2in}
\newcommand{\myShrink}{\hspace*{-0.75in}}
\newcommand{\figHeightThird}{.8in}

\section{Ranking Competitions}
\label{sec:data}
Our next goal is to empirically analyze ranking competitions in which
publishers (document authors) modify documents with respect to
multiple queries ($m > 1$). To the best of our knowledge,
there are no datasets that support this type of
analysis. There are two publicly available datasets~\cite{Raifer+al:17a, Goren+al:20} that resulted from ranking competitions. However, in these competitions, documents were modified with respect to a single query. Thus, we created a new dataset by organizing our
own ranking competitions inspired by those organized in
~\cite{Raifer+al:17a, Goren+al:20}. International and institutional
ethics committees approved the competitions. The participants signed a
consent form and could withdraw at any time.

In addition to having publishers compete for multiple queries rather than a
single one, we introduced two novel aspects in our
competitions. The first is the ranking function. While in previous competitions 
only a feature-based learning-to-rank (\firstmention{\LTR}) method was used~\cite{Raifer+al:17a,Goren+al:20}, we also
used a neural (\firstmention{\sem}) ranker (BERT-based \cite{Nogueira+Cho:19}). The ranking functions are described below. The second
aspect is the use of \llm
tools to modify documents in some of the competitions. In total, we organized four ranking
competitions which differ by the ranking function applied and/or whether generative AI tools were allowed: \firstmention{\A}, \firstmention{\B}, \firstmention{\C},
and \firstmention{\D}. Table~\ref{tab:competition groups} provides an
overview of the competitions.

\begin{table}
\centering
\caption{\label{tab:competition groups} Overview of the four ranking competitions.}
\scriptsize
\begin{tabular}{@{}lcc@{}}
    \toprule
    Competition & Ranking Function & \llm Tools \\
    \midrule
    \A & \lambdamart & \cmark \\
    \B & \lambdamart & \xmark \\
    \C & \bert & \cmark \\
    \D & \bert & \xmark \\
    \bottomrule
\end{tabular}
\end{table}

Each competition lasted for ten rounds; in each round, $30$ games were held, each with respect to a different backstory (topic) from the UQV dataset~\cite{Bailey+al:16a}.\footnote{The dataset includes query variants for TREC's $2013$-$2014$ Web track topics ($201$-$300$).}  Each topic was represented using $m = 3$ queries: the query selected to be the focus of the backstory and two additional query variants selected from all those provided for the backstory. The same $30$ topics were used in all four competitions. Refer to Appendix~\ref{sec:initial} for further details about the topic and query selection procedures.

Before the start of the game, the participants were provided with
three queries representing the topic of the game, the topic description (backstory) and 
an example of a relevant document. This document was generated using
GPT-3.5, as described in Appendix~\ref{sec:initial}. Given the
provided information, the participants were asked to write and submit
their own document. In each round of the game, the participants were
presented with the documents submitted in the previous round and the
rankings induced over these documents for each of the three
queries. They were encouraged to further modify their document to
improve its ranking in the next round. All the documents were in plain
text format, limited to $150$ words. Refer to
Appendix~\ref{sec:instructions} for detailed competition guidelines.

A group of $84$ undergraduate and graduate students enrolled in an information retrieval course participated in our competitions. Each student participated in three (repeated) games and submitted one document per game in a round. We made sure that the three games were selected from at least two different competitions and that there was no overlap between the topics of these games. Each game included exactly five participants: a combination of students and bots. Two or three students participated in every game together with two bots that used GPT-3.5 to create documents, as explained in Appendix~\ref{sec:bots}. If a game had only four participants, i.e., two students and two bots, we added a third static bot that published the initial example document in every round. To prevent any potential bias, the students' identities were anonymized, and they were not informed about the use of bots. Unless otherwise mentioned, our analysis is performed for the students' documents only.

To encourage students to participate and modify their documents so as
to attain high rankings in the competitions, we offered bonus points
for their course grades. We note that the students could have earned a
perfect grade in the course without participating in the
competitions. The bonus per game was awarded based on the median
reciprocal rank of a student's document across the three queries in
the game. For example, if a student's document was ranked first,
second, and third across the queries, they would earn half a point for
that game. The points were accumulated over the student's three games
per round and over the ten rounds.

As noted above, we experimented with two ranking functions in our competitions: \LTR and \sem.
For our learning-to-rank (\LTR) approach, as in previous work~\cite{Raifer+al:17a,Goren+al:20,Vasilisky+al:23}, we use LambdaMART~\cite{ Wu+al:10a} via the RankLib toolkit\footnote{\url{https://sourceforge.net/p/lemur/wiki/RankLib/}}.
Unless otherwise specified, the implementation details of this model, including the training procedure and setting of hyper-parameter values, are the same as those of \citet{Vasilisky+al:23}. The model was trained on the Combined dataset \cite{Vasilisky+al:23}, which includes data from the two previous ranking competitions~\cite{Raifer+al:17a, Goren+al:20}. Each document was represented using a nine-dimensional feature vector. Some of the features are estimates of the similarity between the query and document, and the rest are query-independent document quality measures shown to be effective for Web retrieval~\cite{Bendersky+al:11a}. The features are:
(i) \firstmention{\tf}: sum of query term frequencies in the document,
(ii) \firstmention{\normtf}: \tf normalized by the document's length, 
(iii) \firstmention{\bm}: BM25 similarity between the query and document,
(iv) \firstmention{\lmir}: language-model-based similarity between the query and document\footnote{Unigram language models with Dirichlet smoothing ($\mu = 1000$) are used.}, 
(v) \firstmention{\bert}: BERT-based similarity between the query and document\footnote{We used BERT-Large fine-tuned for passage ranking on MS MARCO~\cite{Nogueira+Cho:19}.},
(vi) \firstmention{\len}: document's length,
(vii) \firstmention{\fracstop}: percentage of terms in the document that are stopwords\footnote{The NLTK stopword list was used in our experiments: \url{www.nltk.org/nltk_data/}.}, (viii) \firstmention{\stopcover}: percentage of stopwords on a stopword list that appear in the document, and
(ix) \firstmention{\ENT}: entropy of the document's term distribution.
For our neural ranking function (\sem), we used \bert as in feature (v) above. Thus, \bert served both as a feature in \LTR and as a standalone ranking function. We note that the ranking functions were not disclosed to the students, who also did not know that four ranking competitions were held.

Our dataset includes a total of $2520$ documents submitted by students ($847$ of which are unique), $2400$ documents generated by bots ($1833$ of which are unique), and $30$ documents that served as the initial document per topic.
We will make the dataset publicly available upon the publication of this work.

Each student document was judged by three crowd workers in
CloudResearch's connect platform \cite{cloudresearch:23a} for binary relevance to the topic and by five workers for content
quality, with $0.75$ and $0.49$ inter-annotator agreement rates (free-marginal multi-rater Kappa), respectively. All workers were native English speakers. More than $97\%$ of all student documents were judged relevant by at least
two out of three annotators. This high relevance percentage is in accordance with
previous reports on competitions held for a single
query \cite{Raifer+al:17a,Goren+al:21a}, albeit a bit
elevated. Although students were not instructed to write relevant
documents, the documents were short and students modified them for
multiple queries representing the same topic. We note that
our focus in this paper is on ranking-incentivized manipulations and not on ranking effectiveness.

Five annotators judged each student document for content quality using
the categories: valid, keyword stuffed and spam (useless). When using
AI tools, $89\%$ and $71\%$ of the documents were judged to be valid
by at least three and at least four out of the five annotators,
respectively. When not using AI tools, the percentages were $85\%$ and
$61\%$, respectively.  This attests to the merit of using AI tools for
rank promotion while maintaining content quality. Only $9\%$ and $6\%$
of the documents created with no AI tools and with AI tools,
respectively, were judged as keyword stuffed by at least three
annotators. Almost none of the documents were judged as spam. The
distribution of quality judgments was not different between the \LTR and \sem competitions.



\section{Empirical Analysis}
\label{sec:eval}
\label{sec:analysis}
\subsection{Analysis of Strategies}
\label{sec:strategies}

\begin{figure*}
    \centering
    \begin{subfigure}{0.25\linewidth}
        \includegraphics[width=\linewidth]{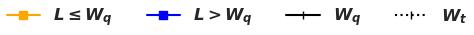}
    \end{subfigure}
   
    \input{figures4/query_independent}
    \hfill
    \input{figures4/query_dependent}
\caption{Averaged feature values of documents that lost in at least four consecutive rounds before winning. The documents are grouped based on whether their feature values four rounds before the win were lower than or equal to (\llewq) or higher than (\lgwq) the values of the winners for the given query. \wq and \wt: the averaged feature values of the corresponding winner per query and all the winners per the same topic, respectively. x-axis: the (negative) number of rounds leading up to a win. Note: in \C and \D, when \bert is used as a standalone ranking function, the feature values of the losing documents are always lower than those of the winning document; therefore, \lgwq is not shown.}
\label{fig:all-features}
\end{figure*}

\myparagraph{Feature Values}
Inspired by work on the single-query setting~\cite{Raifer+al:17a}, we
examine the document modification strategies in our multiple-queries
setting: we analyze the changes in the documents' feature values along
competition rounds. Recall that three document rankings were induced
per topic, each for one of the three queries representing the
topic. In this analysis, we examine each query (document ranking)
separately.  We focus on changes in documents that lost (i.e., were
not ranked first) for at least four consecutive rounds before reaching
the first rank ($\mathbf{L}$).  We compare the average feature values
of the winners (i.e., the highest ranked documents) per query (\wqf)
with the average feature values of two types of losing documents
($L$): those whose feature value was lower than or equal to that of
the winning document four rounds prior to their win (\llewqf), and
those whose feature value was higher (\lgwqf).  We also present for
reference the average feature values of the (at most three) winning
documents per topic (\wtf).  Figure~\ref{fig:all-features} presents
the results for representative query-independent and
query-dependent features of \LTR. Figures for other features exhibit
the same patterns and are omitted as they convey no additional insight.



We see in Figure \ref{fig:all-features} that regardless of the ranking
function (\LTR or \sem), the use of generative \llm tools (\llm or
$\urcorner$\llm) and the initial feature values (\llewq or \lgwq),
with a few exceptions, the features of the losing documents tend to
gradually converge to those of the winners. This finding is in line
with results for the setting where publishers modified their documents
for a single query ~\cite{Raifer+al:17a}. Contrary to
~\citet{Raifer+al:17a}, we see that the feature values of the winning
documents changed throughout the competition rounds. This might be
attributed to the fact that documents in our competitions were
modified for three different queries. For example, say a document was
ranked first for some query. The student who wrote the document may
have continued to modify it to improve its ranking for the other two
queries, and thus affected its feature values. We delve deeper into this point in
Section~\ref{subsec:ranking-functions}.

We also note that changes of winners' feature values conceptually echo our game theoretic findings in Section
\ref{sec:model}: in the single-query setting an equilibrium is
guaranteed in contrast to the multiple-queries setting. Hence, the
single-query setting is more likely to reach a stable state than the multiple-queries setting\footnote{We hasten to point out that the game theoretic analysis was performed for single peak functions. The rankers used in the competition are not single peaked.}

We observe in
Figure~\ref{fig:query-independent} a general upward
trend for the query-independent features: \len, \stopcover, and \ENT; the trends are more pronounced for \A and \B. Hence, the 
documents
became longer (\len) with increased content diversity (\ENT) and
stopwords occurrence (\stopcover).  Interestingly,
\citet{Raifer+al:17a} found a downward trend for \ENT when students
modified documents for a single query. We therefore hypothesize that
when competing for a single query there is an attempt to focus the
content for this query, while competing for multiple queries results
in increased diversity so as to better cover the query set. 



We see in Figure~\ref{fig:query-dependent} that the query-dependent
feature values for \llewq (where the initial feature value of the
losing document was lower than or equal to that of the winner)
increased along the rounds, whereas the values for \lgwq (where the
initial feature value of the losing document was higher than that of
the winner) somewhat fluctuated and even decreased at times. This
implies that in the former case, the students gradually increased the
number of query term occurrences in their documents, but in the latter
case, the students were careful not to add too many query terms,
possibly to avoid keyword stuffing which they were warned
about. Furthermore, this finding potentially suggests that from the point of view
of students, the ranking function might be single peaked with
respect to query term occurrence and the features based on it.

Finally, we can see that the differences between \wq and \wt are smaller for \ltr than for \sem. This finding might be attributed to the fact that for \ltr, a winning document was more likely to win for multiple queries, resulting in fewer unique winning documents per topic. We present further support to this claim in Section~\ref{subsec:ranking-functions}.

\myparagraph{Prompts} To gain additional insights about the document
modification strategies, we asked the students to share the prompts
they used for the generative \llm tools. We collected $44$ prompts
from rounds $8$ to $10$ of the competitions.  The prompts typically
included instructions to create a document that would be ranked high
for the specified queries. All prompts, except for three, had no mention of the reward mechanism used in the competitions.
As an example, one of the three prompts included the text: ``Since the score is
given by the median ranking among 3 queries, we can focus on
maximizing the rankings on two queries.''

Out of the $44$ prompts, $29$ included example documents: the
initial example document ($2$ prompts), the student's document from
the previous round ($15$ prompts), at least one of the winning
documents from the previous round ($24$ prompts), or even all the
documents in the previous round ($5$ prompts). This further indicates
that the students considered winning documents a valuable source
of information about the undisclosed ranking functions.

\subsection{Ranking Functions}
\label{subsec:ranking-functions}

\begin{figure}[t]
\center
\includegraphics[width=\figWidth,height=\figHeight]{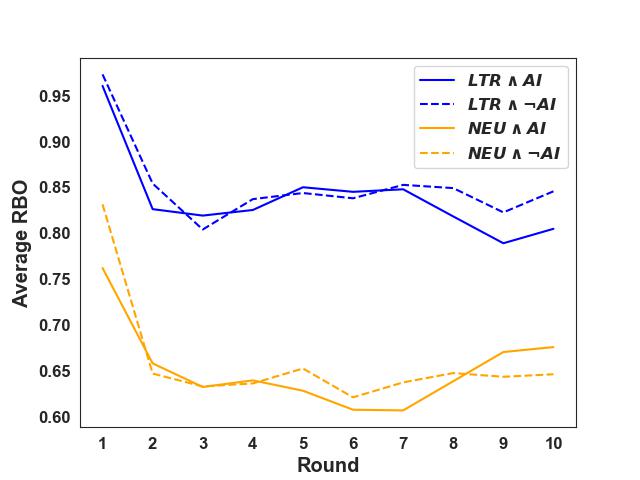}
\caption{\label{fig:rbo} Average RBO similarity between all pairs of rankings induced per topic, averaged over topics per round.}
\end{figure}

Thus far, we studied the document modification strategies in our
competitions. We now shift our focus to examining the effect of
the ranking functions used. We first analyze the differences between
rankings induced by a function for the three queries
representing a topic. We computed for each
ranking function the RBO similarity (with
$p=0.9$)~\cite{Webber+al:2010a} between the document rankings induced
for each pair of queries per topic. We then averaged these values over
the pairs per topic and over topics. Figure~\ref{fig:rbo} presents the
results for each round of the four competitions.  We see that the
similarities for \C and \D are considerably and consistently lower
than those of \A and \B, suggesting that \sem (BERT) is much more
sensitive to the queries used than \ltr (LambdaMART). Previous
research demonstrated BERT's sensitivity to document modifications in
competitions held for a single query~\cite{Vasilisky+al:23}.

We further study the effect of the ranking functions on the
competitions in Table~\ref{tab:num-wins-per-doc} by examining
documents that won for at least one of the queries per topic. We
present the percentage of these documents that won for exactly one,
two or three queries.  Additionally, we present the average ranks of
these documents in rankings induced for queries for which these
documents did not win.  We can see that for \ltr, at least $65\%$ of
the winners won all three queries in a game, while for \sem, less than
$9\%$ did so. Furthermore, the average rank of documents for
queries they did not win was, on average, higher when \sem was used
than when \ltr was used.\footnote{The rank of the highest-ranked document
  is $1$. Increased rank means lower ranking.} These findings demonstrate that the ranking function used
affects the ability of a document to win for multiple queries per
topic. That is, a document that won for one of the queries is more
likely to win for additional queries for the same topic in
competitions that employed \ltr compared to those that employed \sem.

\begin{table}[t]
\centering
\caption{\label{tab:num-wins-per-doc} \docsPercentage: percentage of winning documents that won for exactly $x$ queries per topic. \avgRank: for documents that won for $x$ queries per topic, the average ranks with respect to the $3-x$ queries for which they did not win.}
\scriptsize
\begin{tabular}{@{}lllllllll@{}}
\toprule
 & \multicolumn{2}{c}{\A} & \multicolumn{2}{c}{\B} & \multicolumn{2}{c}{\C} & \multicolumn{2}{c}{\D} \\
 & \docsPercentage & \avgRank & \docsPercentage & \avgRank & \docsPercentage & \avgRank & \docsPercentage & \avgRank \\
\midrule
\OneWin & 17.9\% & 2.2 & 17.7\% & 2.4 & 62.9\% & 3.4 & 65.7\% & 3.5 \\
\TwoWins & 17.0\% & 2.1 & 15.2\% & 2.3 & 28.5\% & 3.3 & 27.6\% & 3.2 \\
\ThreeWins & 65.1\% & - & 67.0\% & - & 8.6\% & - & 6.7\% & - \\
\bottomrule
\end{tabular}

\end{table}

\subsection{Generative \llm Tools}
\label{subsec:llm}

\begin{figure}[t]
\center
\includegraphics[width=\figWidth,height=\figHeight]{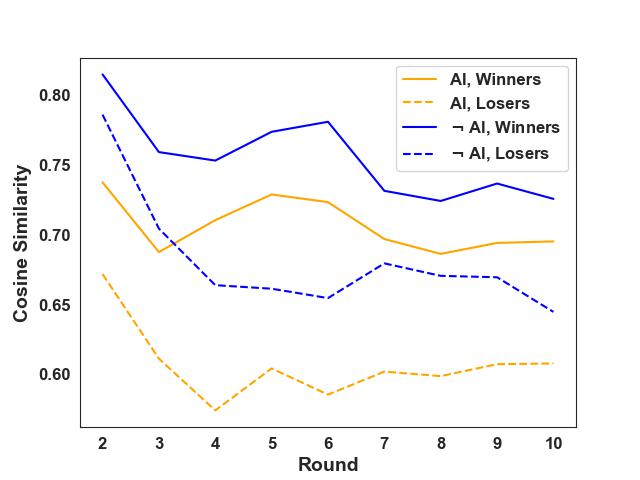}
\caption{\label{fig:originality} Average similarity between {\em all} documents
  in a round {\em per topic} and (i) Winners: documents that won at
  least one query for the topic in the previous round and (ii) Losers:
  all other documents in the previous round.}
\end{figure}

We allowed the use of generative \llm tools in some of our
competitions. We next compare the documents
created by students who were allowed to use such tools (\llm) with
those who were not allowed (\nollm).  We use the cosine of tf.idf
vectors to compute per topic the similarity between each document
submitted by a student in a specific round and (i) documents in the
previous round that won for at least one of the queries and (ii) all
the other documents from the previous round. We averaged the
similarities over the documents and grouped them by whether or not the
students were allowed to use \llm tools. We see in
Figure~\ref{fig:originality} that there were noticeable differences in
the similarity values of \llm and \nollm. Specifically, the documents
created without the help of \llm tools in \B and \D were much more
similar to documents submitted in the previous round, whether they
were winners or not, than those created with the help of such tools in
\A and \C. This finding indicates that using \llm tools may lead to
the creation of more diverse content. We also observe higher
similarities with the winning documents than with the losing ones for
both \llm and \nollm, further supporting our earlier findings that the
documents, whether they were directly modified by students or
generated using \llm tools, tended to converge to the winners.

\section{Predicting Winners}
\label{sec:prediction}
In this section, we address the challenge of predicting which document
among all those that did not rank first in round $\round-1$ will be
ranked the highest (win) in round $\round$ if the winner indeed changes.  The documents in our
competitions were modified with respect to $m=3$ queries in the query set $Q$ for a topic; yet, the prediction is performed
separately for each query $\query \in Q$. A similar challenge was previously tackled for competitions held for a single
query~\cite{Raifer+al:17a}.

For prediction, we represent each document as a $27$-dimensional
feature vector. The features we use can be divided into seven sets. The first two sets, \firstmention{\atomicFt} ($12$ features)
and \firstmention{\metaFt} ($3$ features), were proposed
by~\citet{Raifer+al:17a}.  These features model properties of the
document in round $\round$ ($\docr{\round}$), the document published
by the same author in round $\round-1$ ($\docr{\round-1}$), and the
winning document in round $\round-1$ for the query $\query$
($\winr{\round-1}$).  The features in \atomicFt quantify the changes
in $\docr{\round}$ with respect to $\winr{\round-1}$ by counting term
additions and deletions. These features are defined for three groups
of unique terms: query terms, stopwords, and terms that are neither
query terms nor stopwords. For each group of terms, four features were
defined: the number of terms in $\winr{\round-1}$ that were either (i)
added to or (ii) deleted from $\docr{\round-1}$, and the number
of terms that were not in $\winr{\round-1}$ and were either (iii)
deleted from or (iv) added to $\docr{\round-1}$.  The features
in \metaFt are the similarities between $\docr{\round}$,
$\docr{\round-1}$, and $\winr{\round-1}$; the similarity between a
pair of documents was computed using the cosine of their tf.idf vector
representations.


The features just discussed only consider the winner for the given query ($\winr{\round-1}$), disregarding the rankings induced for the other two queries and their corresponding winners.
Accordingly, we introduce an additional set of features, \firstmention{\centroidMacro}, that consider the group of all winners for the three queries in $Q$ ($\allwinr{\round-1}$). We compute the similarities between $\docr{\round}$ and $\allwinr{\round-1}$ and between $\docr{\round-1}$ and $\allwinr{\round-1}$.
We use cosine to compute the similarities and represent the documents in $\allwinr{\round-1}$ via the centroid of their tf.idf vectors.

In the following two feature sets, we quantify properties of
$\docr{\round-1}$\text{'s} ranks in the three rankings induced for all three
queries per topic. The first set, \firstmention{\PrevPosition},
includes three features: the minimum, median, and maximum rank of
$\docr{\round-1}$ across the three queries in $Q$. The second
set, \firstmention{\QueryPrev}, includes three binary features that
are set to $1$ if $\docr{\round-1}$'s minimum, median, and maximum
rank for the three queries is equal to its rank for the given query
$\query$; otherwise, they are set to $0$.

The next feature set, \firstmention{\PrevImprove}, includes three features that quantify past changes (improvements) in the document's ranks. The underlying assumption is that if a publisher improved her document's rank in round $\round-1$, she might do so again in round $\round$. To this end, we also examine the publisher's document in round $\round-2$ ($\docr{\round-2}$). For each of the three queries, we compute the difference between the ranks of $\docr{\round-1}$ and $\docr{\round-2}$, and scale it by the maximal possible rank change given $\docr{\round-2}$'s rank in round $\round-2$. We use the minimum, median, and maximum values across the three queries.

Finally, we found in Section~\ref{sec:strategies} that the documents tended to become longer throughout the four competition rounds leading to a win; accordingly, the feature in the final set, \firstmention{\docLen}, is $\docr{\round}$'s length.

\subsection{Setup}
\label{sec:expSetup}

Our dataset includes documents submitted by both students and bots. We discarded queries for which a bot generated the winning document. In addition, we discarded queries for which the winners in round $\round-1$ ($\winr{\round-1}$) and round $\round$ ($\winr{\round}$) were generated by the same publisher. This was done because we aimed to predict which of the losing documents would win in the following round. As a result, the number of queries per competition round ranged from $3$ to $39$.

We experimented with four different classifiers with the features defined above via the scikit-learn library~\cite{scikit-learn}: logistic regression (\firstmention{\lreg}), linear SVM (\firstmention{\lsvm}), polynomial SVM (\firstmention{\psvm}) and random forests (\firstmention{\rforest}). We applied min-max normalization to feature values per query.
The document assigned the highest prediction score per query by a classifier was deemed the winner; the remaining documents were considered losers.

The features in \PrevImprove can only be computed when data for at least two previous rounds is available; therefore, our experiments were conducted for rounds $3$ to $10$. We used leave-one-out cross-validation over rounds to train our models and select parameter values. 
We used the queries from one of the rounds for testing and those from the remaining seven for training and validation. We repeatedly trained a model on data from six of the seven rounds with different parameter configurations and validated the effectiveness over the seventh round. We selected the parameter values that yielded the highest prediction effectiveness on average across the seven validation rounds. Then, we trained a final model using data from all seven rounds and applied it to the held-out test round (fold). We repeated this entire procedure for each test fold. The models were trained separately for each of the four competitions.

We use \firstmention{\acc} (percentage of correctly classified winners and losers) and \firstmention{\FOne} (harmonic mean of precision and recall), averaged over queries and test folds, to measure prediction effectiveness; the former served to select parameter values.
Statistically significant prediction differences were determined across queries using the two-tailed paired t-tests with $p\leq0.05$.
\lreg, \lsvm and \psvm were trained with L1 regularization; the regularization parameter was selected from~$\set{1,10,50,100}$. The degree of the polynomial kernel in \psvm was selected from~$\set{2,3,4,5}$. For \rforest, the number of trees and leaves were set to values in~$\set{10,50,100,500}$ and~$\set{10,20,30}$, respectively. All other parameters were set to default values~\cite{scikit-learn}.

\begin{table}[t]
  \setlength{\tabcolsep}{3pt}
\caption{\label{tab:main} Prediction effectiveness. All differences between our classifiers (\lreg, \lsvm, \psvm, \rforest) and the baselines are statistically significant. Bold: best result in a row.}
\scriptsize
\begin{tabular}{@{}llllllllllll@{}}
\toprule
  &   & \randomShort & \qmajorityShort & \tmajorityShort & \awinnerShort & \aloserShort & \nimrodBestShort & \lreg & \lsvm & \psvm & \rforest \\
\midrule
\A & \acc & 0.654 & 0.435 & 0.503 & 0.466 & 0.534 & 0.696 & 0.822 & 0.853 & \textbf{0.895} & 0.843 \\
\A & \FOne & 0.622 & 0.397 & 0.497 & 0.635 & 0.000 & 0.678 & 0.810 & 0.843 & \textbf{0.885} & 0.832 \\ \midrule
\B & \acc & 0.707 & 0.398 & 0.504 & 0.480 & 0.520 & 0.772 & \textbf{0.886} & 0.870 & 0.846 & 0.837 \\
\B & \FOne & 0.697 & 0.368 & 0.497 & 0.648 & 0.000 & 0.763 & \textbf{0.879} & 0.863 & 0.839 & 0.831 \\ \midrule
\C & \acc & 0.669 & 0.532 & 0.468 & 0.464 & 0.536 & 0.766 & 0.815 & 0.798 & 0.806 & \textbf{0.855} \\
\C & \FOne & 0.643 & 0.495 & 0.588 & 0.634 & 0.000 & 0.747 & 0.800 & 0.782 & 0.790 & \textbf{0.842} \\ \midrule
\D & \acc & 0.556 & 0.481 & 0.500 & 0.491 & 0.509 & 0.722 & 0.750 & 0.750 & \textbf{0.796} & 0.750 \\
\D & \FOne & 0.547 & 0.471 & 0.592 & 0.658 & 0.000 & 0.717 & 0.742 & 0.742 & \textbf{0.790} & 0.743 \\
\bottomrule
\end{tabular}

\end{table}

\subsection{Results}
\label{sec:results}

We compare in Table~\ref{tab:main} the prediction effectiveness of our
four classifiers (\lreg, \lsvm, \psvm, and \rforest) with that of five
baselines\footnote{Except for \tmajority, all the baselines were also
used by~\citet{Raifer+al:17a}.}: (i) {\random} (\firstmention{\randomShort}) : the
winner is randomly selected, (ii) {\qmajority} (\firstmention{\qmajorityShort}) : the
winner is the document whose publisher had the highest number of past
wins for the given query; ties were broken randomly,
(iii) {\tmajority} (\firstmention{\tmajorityShort}): the winner is the document whose
publisher had the highest number of past wins for all three queries of
the given topic; ties were broken randomly,
(iv) {\awinner} (\firstmention{\awinnerShort}): all the documents are predicted to be
winners (only one document is correctly classified per query),
(v) {\aloser} (\firstmention{\aloserShort}): all the documents are predicted to be
losers (all but one of the documents are correctly classified per
query). We also report the effectiveness of the classifiers when using
only the two feature sets proposed by~\citet{Raifer+al:17a} (\atomicFt
and \metaFt). For this baseline, we show the results only for the 
classifier (\lreg, \lsvm, \psvm, or \rforest) with the highest \acc per competition.

The documents in our competitions were modified with respect to three
queries.  We see in Table~\ref{tab:main} that when using only the two
feature sets proposed by~\citet{Raifer+al:17a}, which use information
only about the query at hand, the performance surpasses that of all
baselines. This attests to the effectiveness of these feature sets,
whether the documents are modified with respect to a single query as
in~\cite{Raifer+al:17a} or with respect to multiple queries as in our
setting.  Furthermore, we see that when adding features that quantify
various properties of all three queries and the corresponding
rankings, all our classifiers consistently and statistically
significantly outperform all the baselines, including that
of~\citet{Raifer+al:17a}. This attests to the merits of using
information about other queries for the same topic to predict for a given query.

%

To further examine the effectiveness of the different feature sets, we performed ablation tests, removing one of the sets each time. We anzlye the results for \psvm, for which the best overall performance was attained for two of the four competitions in Table~\ref{tab:main}. Actual numbers are omitted as they convey no additional insight and due to space considerations.
Except for \metaFt, removing each feature set results in at least one statistically significant drop in performance, attesting to the complementary nature of our features. Comparing \metaFt and \centroidMacro, the drop for the latter is more substantial. These findings suggest that these two feature sets might be somewhat correlated and that \centroidMacro, which considers the winners for all queries of the topic in the previous round, is somewhat more informative in our setting.


\omt{

\begin{table*}[t]
\centering
\footnotesize
\caption{\label{tab:main} Single }
\begin{tabular}{lllllllll}
\toprule
  & \multicolumn{2}{c}{\A} & \multicolumn{2}{c}{\B} & \multicolumn{2}{c}{\C} & \multicolumn{2}{c}{\D} \\
  & \acc & \FOne & \acc & \FOne & \acc & \FOne & \acc & \FOne \\
\midrule
\atomicFt & $0.613^*$ & $0.588^*$ & $0.634^*$ & $0.617^*$ & $0.742$ & $0.720$ & $0.602^*$ & $0.593^*$ \\
\metaFt & $0.665^*$ & $0.642^*$ & $0.667^*$ & $0.654^*$ & $0.750$ & $0.729$ & $0.639^*$ & $0.633^*$ \\
\centroidMacro & $0.696^*$ & $0.667^*$ & $0.577^*$ & $0.562^*$ & $0.589^*$ & $0.554^*$ & $0.463^*$ & $0.455^*$ \\
\docLen & $0.822$ & $0.809$ & $0.740^*$ & $0.731^*$ & $0.468^*$ & $0.423^*$ & $0.556^*$ & $0.551^*$ \\
\PrevPosition & $0.581^*$ & $0.554^*$ & $0.537^*$ & $0.515^*$ & $0.524^*$ & $0.489^*$ & $0.583^*$ & $0.575^*$ \\
\QueryPrev & $0.550^*$ & $0.508^*$ & $0.593^*$ & $0.580^*$ & $0.500^*$ & $0.458^*$ & $0.509^*$ & $0.502^*$ \\
\PrevImprove & $0.560^*$ & $0.528^*$ & $0.480^*$ & $0.452^*$ & $0.556^*$ & $0.521^*$ & $0.519^*$ & $0.509^*$ \\
\allPredFeatures & $\mathbf{0.895}$ & $\mathbf{0.885}$ & $\mathbf{0.846}$ & $\mathbf{0.839}$ & $\mathbf{0.806}$ & $\mathbf{0.790}$ & $\mathbf{0.796}$ & $\mathbf{0.790}$ \\
\bottomrule
\end{tabular}

\end{table*}

}

\omt{

\paragraph*{Baselines}

\begin{table*}[t]
\centering
\footnotesize
\caption{\label{tab:baselines} Prediction effectiveness of the baselines and the four classifiers (\lreg, \lsvm, \psvm, \ourAlg) trained with the original feature sets: \atomicFt and \metaFt.
Results are given for each of the groups (A-D) and for \nimrodComp.
Excluding the \random baseline, the difference between each classifier and each baseline is statistically significant.
Bold: the best result in a row.}

\begin{tabular}{llrrrrrrrrr}
  \toprule
 &  & \random & \qmajority & \tmajority & \awinner & \aloser & \lreg & \lsvm & \psvm & \rforest \\
 \midrule
\A & \acc & 0.647 & 0.395 & 0.504 & 0.466 & 0.534 & 0.714 & 0.711 & 0.746 & \textbf{0.757} \\
\A & \FOne & 0.622 & 0.351 & 0.491 & 0.636 & 0.0 & 0.698 & 0.694 & 0.728 & \textbf{0.743} \\
  \midrule
\B & \acc & 0.758 & 0.406 & 0.506 & 0.476 & 0.524 & 0.721 & 0.708 & 0.699 & \textbf{0.791} \\
\B & \FOne & 0.745 & 0.376 & 0.479 & 0.645 & 0.0 & 0.708 & 0.694 & 0.685 & \textbf{0.783} \\
  \midrule
\C & \acc & 0.689 & 0.509 & 0.471 & 0.464 & 0.536 & \textbf{0.722} & 0.715 & 0.669 & 0.713 \\
\C & \FOne & 0.664 & 0.47 & 0.583 & 0.634 & 0.0 & \textbf{0.701} & 0.694 & 0.645 & 0.692 \\
  \midrule
\D & \acc & 0.648 & 0.461 & 0.502 & 0.487 & 0.513 & 0.733 & \textbf{0.742} & 0.642 & 0.72 \\
\D & \FOne & 0.638 & 0.446 & 0.588 & 0.655 & 0.0 & 0.727 & \textbf{0.735} & 0.635 & 0.712 \\
  \midrule
  \midrule
\nimrodComp & \acc & 0.627 & 0.685 & - & 0.247 & 0.753 & 0.849 & 0.859 & 0.867 & \textbf{0.878} \\
\nimrodComp & \FOne & 0.242 & 0.363 & - & 0.396 & 0.0 & 0.695 & 0.712 & 0.730 & \textbf{0.752} \\
 \bottomrule
\end{tabular}

\end{table*}

We use the following baselines for comparison of prediction effectiveness:

\begin{itemize}
    \item \firstmention{\random}: the winner is randomly selected.
    \item \firstmention{\qmajority}: the predicted winner is the document whose publisher had the most past wins for the specific query (in cases of ties, a random choice is made).
    \item \firstmention{\tmajority}: the predicted winner is the document whose publisher had the most past wins for all three queries belonging to the same topic (in cases of ties, a random choice is made).
    \item \firstmention{\awinner}: all documents are predicted to be winners, so only one prediction is correct.
    \item \firstmention{\aloser}: all documents are predicted to be losers, so all but one prediction is correct.
\end{itemize}

Except for \tmajority, all baselines are the same as those used by~\cite{Raifer+al:17a}.
In Table~\ref{tab:baselines} we compare the prediction effectiveness of the four classifiers trained with the original feature sets (\atomicFt and \metaFt) with that of the five baselines.
We additionally include the results of \nimrodComp.
We see that the performance of all classifiers is worse than of the same task but with a single query.
Moreover, while the \ourAlg classifier had the best performance in the single query task, here it is outperformed by \lreg and \lsvm for groups with the \bert ranker (C and D).

\begin{table*}[t]
\centering
\footnotesize
\caption{\label{tab:ablation} Performance with subsets of features for prediction. Bold: best result.}
\subcaption*{\A}
\begin{tabular}{lllllllll}
\toprule
  & \multicolumn{2}{c}{LReg} & \multicolumn{2}{c}{LSVM} & \multicolumn{2}{c}{PSVM} & \multicolumn{2}{c}{RForest} \\
  & \acc & \FOne & \acc & \FOne & \acc & \FOne & \acc & \FOne \\
  &  &  &  &  &  &  &  &  \\
\midrule
\metaFt + \atomicFt & 0.714 & 0.698 & 0.711 & 0.694 & 0.746 & 0.728 & 0.757 & 0.743 \\
\docLen + \metaFt + \atomicFt & 0.835 & 0.824 & 0.829 & 0.817 & 0.882 & 0.872 & 0.896 & 0.889 \\
\centroidMacro + \metaFt + \atomicFt & 0.79 & 0.775 & 0.794 & 0.779 & 0.788 & 0.772 & 0.821 & 0.808 \\
\metaFt + \atomicFt + \QueryPrev & 0.746 & 0.731 & 0.744 & 0.729 & 0.763 & 0.746 & 0.779 & 0.764 \\
\metaFt + \atomicFt + \PrevPosition & 0.752 & 0.736 & 0.745 & 0.729 & 0.831 & 0.819 & 0.835 & 0.823 \\
\metaFt + \atomicFt + \PrevImprove & 0.686 & 0.671 & 0.666 & 0.649 & 0.769 & 0.756 & 0.753 & 0.74 \\
\centroidMacro + \docLen + \metaFt + \atomicFt + \PrevPosition + \QueryPrev + \PrevImprove & 0.858 & 0.847 & 0.87 & 0.858 & 0.891 & 0.881 & 0.843 & 0.832 \\
\docLen + \metaFt + \QueryPrev & \textbf{0.912} & \textbf{0.905} & \textbf{0.912} & \textbf{0.905} & 0.874 & 0.863 & 0.873 & 0.864 \\
\bottomrule
\end{tabular}

\bigskip
\subcaption*{\B}
\begin{tabular}{lllllllll}
\toprule
  & \multicolumn{2}{c}{LReg} & \multicolumn{2}{c}{LSVM} & \multicolumn{2}{c}{PSVM} & \multicolumn{2}{c}{RForest} \\
  & \acc & \FOne & \acc & \FOne & \acc & \FOne & \acc & \FOne \\
  &  &  &  &  &  &  &  &  \\
\midrule
\metaFt + \atomicFt & 0.721 & 0.708 & 0.708 & 0.694 & 0.699 & 0.685 & 0.791 & 0.783 \\
\docLen + \metaFt + \atomicFt & 0.866 & 0.858 & 0.868 & 0.861 & 0.823 & 0.813 & 0.878 & 0.871 \\
\centroidMacro + \metaFt + \atomicFt & 0.738 & 0.727 & 0.733 & 0.721 & 0.722 & 0.708 & 0.785 & 0.777 \\
\metaFt + \atomicFt + \QueryPrev & 0.747 & 0.735 & 0.743 & 0.731 & 0.711 & 0.699 & 0.801 & 0.793 \\
\metaFt + \atomicFt + \PrevPosition & 0.694 & 0.68 & 0.683 & 0.669 & 0.73 & 0.719 & 0.765 & 0.756 \\
\metaFt + \atomicFt + \PrevImprove & 0.756 & 0.74 & 0.763 & 0.749 & 0.696 & 0.682 & 0.792 & 0.782 \\
\centroidMacro + \docLen + \metaFt + \atomicFt + \PrevPosition + \QueryPrev + \PrevImprove & 0.838 & 0.83 & 0.865 & 0.858 & 0.865 & 0.859 & 0.847 & 0.838 \\
\docLen + \atomicFt + \PrevPosition + \QueryPrev + \PrevImprove & 0.909 & 0.903 & \textbf{0.914} & \textbf{0.908} & 0.838 & 0.828 & 0.862 & 0.852 \\
\bottomrule
\end{tabular}

\bigskip
\subcaption*{\C}
\begin{tabular}{lllllllll}
\toprule
  & \multicolumn{2}{c}{LReg} & \multicolumn{2}{c}{LSVM} & \multicolumn{2}{c}{PSVM} & \multicolumn{2}{c}{RForest} \\
  & \acc & \FOne & \acc & \FOne & \acc & \FOne & \acc & \FOne \\
  &  &  &  &  &  &  &  &  \\
\midrule
\metaFt + \atomicFt & 0.722 & 0.701 & 0.715 & 0.694 & 0.669 & 0.645 & 0.713 & 0.692 \\
\docLen + \metaFt + \atomicFt & 0.709 & 0.687 & 0.702 & 0.679 & 0.659 & 0.633 & 0.712 & 0.69 \\
\centroidMacro + \metaFt + \atomicFt & 0.71 & 0.688 & 0.702 & 0.68 & 0.653 & 0.627 & 0.703 & 0.68 \\
\metaFt + \atomicFt + \QueryPrev & 0.728 & 0.707 & 0.725 & 0.704 & 0.697 & 0.672 & 0.709 & 0.687 \\
\metaFt + \atomicFt + \PrevPosition & 0.772 & 0.755 & 0.771 & 0.754 & 0.767 & 0.749 & 0.761 & 0.744 \\
\metaFt + \atomicFt + \PrevImprove & 0.73 & 0.713 & 0.746 & 0.729 & 0.712 & 0.692 & 0.771 & 0.755 \\
\centroidMacro + \docLen + \metaFt + \atomicFt + \PrevPosition + \QueryPrev + \PrevImprove & 0.811 & 0.797 & 0.803 & 0.788 & 0.814 & 0.798 & 0.826 & 0.812 \\
\metaFt + \atomicFt + \PrevPosition + \QueryPrev + \PrevImprove & 0.814 & 0.8 & 0.814 & 0.8 & 0.824 & 0.81 & \textbf{0.838} & \textbf{0.825} \\
\bottomrule
\end{tabular}

\bigskip
\subcaption*{\D}
\begin{tabular}{lllllllll}
\toprule
  & \multicolumn{2}{c}{LReg} & \multicolumn{2}{c}{LSVM} & \multicolumn{2}{c}{PSVM} & \multicolumn{2}{c}{RForest} \\
  & \acc & \FOne & \acc & \FOne & \acc & \FOne & \acc & \FOne \\
  &  &  &  &  &  &  &  &  \\
\midrule
\metaFt + \atomicFt & 0.733 & 0.727 & 0.742 & 0.735 & 0.642 & 0.635 & 0.72 & 0.712 \\
\docLen + \metaFt + \atomicFt & 0.754 & 0.748 & 0.76 & 0.753 & 0.637 & 0.629 & 0.718 & 0.71 \\
\centroidMacro + \metaFt + \atomicFt & 0.752 & 0.745 & 0.762 & 0.755 & 0.656 & 0.648 & 0.712 & 0.704 \\
\metaFt + \atomicFt + \QueryPrev & 0.717 & 0.71 & 0.726 & 0.719 & 0.655 & 0.647 & 0.734 & 0.727 \\
\metaFt + \atomicFt + \PrevPosition & 0.725 & 0.716 & 0.721 & 0.713 & 0.748 & 0.741 & 0.747 & 0.739 \\
\metaFt + \atomicFt + \PrevImprove & 0.717 & 0.712 & 0.715 & 0.709 & 0.661 & 0.655 & 0.73 & 0.725 \\
\centroidMacro + \docLen + \metaFt + \atomicFt + \PrevPosition + \QueryPrev + \PrevImprove & 0.781 & 0.774 & 0.784 & 0.778 & 0.785 & 0.779 & 0.752 & 0.746 \\
\centroidMacro + \metaFt + \atomicFt + \PrevPosition + \PrevImprove & 0.809 & 0.803 & \textbf{0.819} & \textbf{0.814} & 0.701 & 0.696 & 0.751 & 0.744 \\
\bottomrule
\end{tabular}

\end{table*}

\paragraph*{Feature analysis}
In order to improve the prediction effectiveness, we added the features described in Section~\ref{sec:features}.
In Table~\ref{tab:ablation} we compare the prediction effectiveness of training the four classifiers with different combinations of features.
The features we added were able to improve the prediction effectiveness for all the groups.

}

\section{Conclusions}
We presented a novel theoretical and empirical study of the
competitive retrieval setting where document authors
modify documents to improve their ranking for multiple queries.  We
showed using game theory that in contrast to past work on document modifications for a single query, in the
multiple-queries setting there is not necessarily an equilibrium and
characterized the cases when it exists. We organized 
novel ranking competitions where publishers modified documents for multiple queries representing the same
topic. We used both neural and non-neural rankers, and allowed
the use of generative AI tools in some competitions. We found  
that publishers use content from documents highly ranked in the
past, but to a somewhat reduced extent when they apply AI tools. Finally, we
addressed the task of predicting which document will be promoted to the highest rank in the next ranking round. We demonstrated the merits of using
information induced from other queries representing the same topic to this end.

%
%

\appendix
\section{Ranking Competitions}
\label{app:asrc}

\subsection{Topics, Queries, and Initial Documents}
\label{sec:initial}
We selected from the UQV dataset~\cite{Bailey+al:16a} topics with commercial intent that were more likely to incite competition\footnote{The selected topics are $201$, $203$, $204$, $209$, $210$, $211$, $212$, $218$, $226$, $228$, $233$, $235$, $244$, $245$, $246$, $249$, $250$, $252$, $255$, $258$, $261$, $262$, $268$, $272$, $274$, $281$, $283$, $289$, $291$, and $296$.}.
For each topic, three queries were used. The first query was the focus of the backstory, while the other two were selected from all the available query variants as follows. To ensure that the variants were of high quality, we retrieved $1000$ documents per variant from the ClueWeb12 category B corpus using \lmir via the Indri toolkit\footnote{\url{www.lemurproject.org/indri}}. We filtered out variants with average precision (AP) lower than $0.05$. In addition, to ensure that the queries were diverse enough, we computed the Jacquard similarity between the terms of each pair of queries and the RBO similarity (with $p=0.9$)~\cite{Webber+al:2010a} between the corresponding retrieved document lists. These two scores were combined using Reciprocal Rank Fusion (RRF) with $k=60$~\cite{Cormack+al:09a}. Finally, from the remaining query variants, we iteratively selected the following query with the lowest average similarity to the already selected queries.

We provided the students with an example of a relevant document for
each topic. We generated these documents using GPT-3.5 with a
prompt per topic that included the backstory and the three selected
queries. We continued generating the document until it was
deemed relevant by four annotators. We used the same document per
topic across all four competitions, resulting in $30$ documents.

\subsection{Additional Guidelines}
\label{sec:instructions}

The students were instructed to create high quality plain text documents without using any formatting tags, hyperlinks, or special characters. They were advised to avoid using slang or informal language and were warned against keyword stuffing. We mentioned that using such techniques would result in a penalty, although no actual penalty was imposed. Copying parts of other students' documents or Web pages was allowed as long as the document also included original content. Documents that were copied completely from other documents received no reward.
We noticed after the third round that some students were submitting the initial example document without making any modifications. As a result, we prohibited submitting the initial documents starting from the fourth round. Students who still used these documents did not receive points.
We emphasized that the goal was to promote documents in rankings regardless of their relevance to the information need expressed by the three queries.

\subsection{Bots}
\label{sec:bots}
We used bots in our competitions to increase the number of players and to make the games more dynamic. The bots generated documents using GPT-3.5. The prompts instructed the bots to create documents with a maximum length of $150$ terms, but some of the generated documents still exceeded this limit. In such cases, we removed sentences from the end of a document until its length was between $140$ and $150$ terms. If no sentence could be removed without making the document shorter than $140$ terms, we truncated the last sentence to $150$ terms.

We provided our bots with the backstory, the three queries
representing the backstory, and the content and ranks of the bot's
document in the previous round. In addition, we provided information
about other submitted documents that varied across bots based on two
parameters.  The first parameter determined which of the documents
submitted in a round were accessible to the bot. We tested bots that
had access to all submitted documents (including their ranks), the
document with the highest median rank across the three queries
(excluding the actual ranks), and the bot's document only. The second
parameter determined the number of rounds for which the above
information was made available. We tested bots that had information
about the three previous rounds and those that had information about
the previous round only. We used six bot configurations in
our competitions, each corresponding to a different combination of the
two parameters. Each bot was assigned to ten different topics and
competed in all ten rounds of the game.

We found that the documents generated by bots were generally ranked lower than those written by students. The average rank for the bots' documents was $3.2$, while the students' documents had an average rank of $2.6$. Moreover, the bots' documents were less likely to win in our competitions. Specifically, $67.9\%$ of the winning documents were written by students, compared to only $23.5\%$ that were generated by bots.\footnote{The percentages do not add up to $100$ as we excluded from this analysis the static bots that submitted the initial document in all competition rounds.}


\omt{
\subsection{Guidelines}
\label{sec:instructions}

Participants were directed to compose their documents in plain text, avoiding any formatting tags, hyperlinks, and special characters.
They were allowed to refer to the initial document provided for each topic as an example.
Furthermore, it was clarified that the submissions need not directly align with the information need expressed by the three queries.

The students were instructed to write high-quality documents, avoiding slang and informal language.
Their length was limited to a maximum of 150 words.
Participants were cautioned against keyword stuffing, and threatened that employing such techniques would result in a penalty.
However, copying parts of other participants’ documents or web pages was permissible, as long as the overall document maintained originality.
Documents that were copied completely from other documents obtained a score of zero.
Examples of good and bad practices in keyword usage and document cohesion were provided to guide participants.

After the third round, we noticed that some students were not updating their documents.
We therefore decided to prohibit the use of the initial relevant document as a submission since the fourth round.
Student that still used the initial relevant document as a submission did not receive any reward for that round.

As discussed in section~\ref{sec:data}, in some of the competitions the usage of AI tools was prohibited.
In these competitions, students were instructed to write their documents without the use of any AI tools.
In the other competitions, students were allowed to use AI tools, and even encouraged to do so.
From the seventh round, we asked students in competitions in which AI tools were allowed to provide the prompts they used to generate their documents.
}

\omt{
\subsection{Incentives}
\label{sec:reward}

We incentivized students to participate in the competition by offering extra credit points for their exam.
In each round, each student submitted three documents; one for each match they participated in.
In each match, the documents were ranked three times, once for each query variant.
Students earned points based on the inverse of the median rank; for instance, if a student's document ranked first, second, and third across the queries, they would earn $\frac{1}{2}$ points for that match.
A student's total reward was the sum of points from all the matches they participated in, over all ten rounds.

}

\omt{
\subsection{Queries and initial relevant documents}
\label{sec:initial}

In order to simulate the competitive nature of the Web, we chose topics that are likely to have commercial intent.
The 30 topics were selected from the TREC 2013-2014 Web track (topics $201$-$300$).
The topic (or subtopic) titles were used as queries. \footnote{The selected topics were 261, 204, 212-1, 210-3, 218-2, 235-2, 250, 233-4, 283, 255-4, 201-4, 246, 228, 226-1, 211, 258-4, 274-2, 252, 289-1, 281-1, 209-3, 244-2, 268, 249-5, 262, 272-1, 296, 291-7, 245-1, 203.}

We used the query variations provided by the UQV-100 dataset~\cite{Bailey+al:16a} to select two more queries per topic.
For each query variation, we retrieved 1000 documents from the ClueWeb12 category B corpus using query-likelihood with Dirichlet smoothing ($\mu=1000$) using the Indri toolkit \footnote{\url{www.lemurproject.org/indri}}.
Some variations in the dataset were unrelated to the original topic, for example, for the topic "home theater systems", the query "new house" was provided.
Naturally, these queries were unrelated to the original topic, and thus the documents retrieved for them were distinct from the documents retrieved for the other queries in the topic.
Therefore, to ensure the query variations were related to the original topic, we first filtered out queries that gained $AP$ score of less than $0.05$.

While ensuring the query variations were related to the original topic, we also wanted to promote diversity between the queries.
To this end, we computed the RBO ($p=0.9$) and Jacquard scores between the retrieved documents for each pair of queries, and fused these scores using Reciprocal Rank Fusion (RRF) with $k=60$.
Our goal was to choose the variations that had the lowest similarity between their retrieved documents.
Hence, we first chose the query variation with the lowest score with the original query, and then chose the variation with the lowest average score with the queries already chosen.

%

For the purpose of assisting students in understanding what kind of information the queries were essentially seeking, we provided them with an example of a document for each topic.
We used the same document for each topic across the four competitions.
The initial relevant document was generated using ChatGPT 3.5, and the prompt for each topic included the three queries and the backstory from the UQV-100 dataset~\cite{Bailey+al:16a}.
We iteratively generated documents until we obtained one that was judged relevant by all four (\haya{Haya, Niv, Oren and Fiana}) annotators.

}

\omt{

\subsection{Ranking model}
\label{sec:retMethod}

\paragraph*{LambdaMART}
Similar to the previous competitions, we use a \ltr approach to rank the documents.
However, we follow the approach of~\cite{Vasilisky+al:23} and train the model on previous competitions' data.
Specifically, we use their method to combine the data from~\cite{Raifer+al:17a} and~\cite{Goren+al:20} to a single dataset referred to as \emph{Combined}.
All the features used in~\cite{Raifer+al:17a} and~\cite{Goren+al:20} were used in the \ltr model, except for the BERT feature.
The model was trained via the RankLib library\footnote{\url{https://sourceforge.net/p/lemur/wiki/RankLib/}}.

\paragraph*{BERT}
We use a BERT model that was trained for query-based passage ranking on the MS MARCO dataset~\cite{Nogueira+Cho:19}.
In the \lambdamart model, we used the score assigned to the document by the BERT model as a feature.

}

\omt{
\subsection{Bots}
\label{sec:bots}

\begin{table}[h]
\centering
\caption{\label{tab:bots} Bots names and descriptions.}
\small

\begin{tabular}{|c|c|c|}
    \hline
    \textbf{Bot Name} & \textbf{Observability} & \textbf{Memory} \\
    \hline
    \MABOT & \obsA & \markov \\
    \hline
    \MTBOT& \obsT & \markov \\
    \hline
    \MSBOT & \obsS & \markov \\
    \hline
    \NMABOT & \obsA & \nonmarkov \\
    \hline
    \NMTBOT & \obsT & \nonmarkov \\
    \hline
    \NMSBOT & \obsS & \nonmarkov \\
    \hline
\end{tabular}
\end{table}

In addition to the students, we also introduced bots to the competition.
Our bots differ from one another in their observability and memory levels.

\textbf{Observability level} refers to an agent's ability to perceive its environment.
In this context, the environment is the competing documents.
The observability level of our bots ranges from \obsA (awareness of all documents), to \obsT (visibility of only the top document each round), and \obsS (awareness only of its own document).
\textbf{Memory level} in AI represents an agent's capability to recall past states or actions.
A \nonmarkov memory agent retains information of all previous rounds, while a \markov agent retains only the most recent round information.

Overall, we introduced six bots to the competition, each with a different combination of observability and memory levels.
Table~\ref{tab:bots} summarizes the bots' names and descriptions.

We used the API of GPT-3.5 \footnote{\url{https://openai.com/}}, which was the most advanced version available at the time of the competition, to generate the bots' documents.
\footnote{Often, the documents generated by the bots were too long, even though we limited the length to 150 words in the prompt.
In these cases, we truncated the last sentences of the documents to fit the length limit.}

}

\balance
\bibliography{cunlp-ir,fiana-bib}

\end{document}